\documentclass[journal]{IEEEtran}

\usepackage{xcolor}
\usepackage[normalem]{ulem}
\usepackage{graphicx}
\usepackage{draftwatermark}
\SetWatermarkText{Confidential: For review of PaRTAA}
\SetWatermarkLightness{1}
\SetWatermarkScale{.4}
\usepackage{amsmath}
\usepackage{lineno,hyperref}
\usepackage{rotating, graphicx}
\ifCLASSINFOpdf
\else
\fi

\newcommand{\rev}[1]{{\color{black}  #1}}
\newcommand{\new}[1]{{\color{black}  #1}}
\newcommand{\revtwo}[1]{{\color{black}  #1}}

\begin{document}
\bstctlcite{IEEEexample:BSTcontrol}
\title{\AE r\o : A Platform Architecture for Mixed-Criticality Airborne Systems}
%
%
%

\author{Shibarchi~Majumder~\IEEEmembership{Member,~IEEE,}
           ~Jens~F~Dalsgaard Nielsen and~Thomas~Bak~\IEEEmembership{Senior Member,~IEEE,}
\thanks{The authors are with the Department of Electronic Systems, Aalborg University, Aalborg 9220, Denmark. email: sm, jdn, tba (@es.aau.dk).}
\thanks{This research is funded by Danish Independent Research Foundation under grant number 6111-00363B.}
\thanks{Manuscript received on March 01st 2019; revised on July 12th 2019 and October 09th 2019; Accepted on 22nd November 2019.}}

%
%

\markboth{Published: IEEE Transactions on Computer-Aided Design of Integrated Circuits and Systems,~Vol.~39, Issue.~10, December~2019}%
{Shell \MakeLowercase{\textit{et al.}}: }
%



\maketitle

\begin{abstract}
{

Real-time embedded platforms with resource constraints can take the benefits of mixed-criticality system where applications with different criticality-level share computational resources, with isolation in the temporal and spatial domain. 
A conventional software-based isolation mechanism adds additional overhead and requires certification with the highest level of criticality present in the system, which is often an expensive process. 

In this work, we present a different approach where the required isolation is established at the hardware-level by featuring partitions within the processor.   
A 4-stage pipelined soft-processor with replicated resources in the data-path is introduced to establish isolation and avert interference between the partitions. A cycle-accurate scheduling mechanism is implemented in the hardware for hard-real-time partition scheduling that can accommodate different periodicity and execution time for each partition as per user needs, while preserving time-predictability at the individual application level. Applications running within a partition has no sense of the virtualization and can execute either on a host-software or directly on the hardware. The proposed architecture is implemented on FPGA thread and demonstrated with an avionics use case.}

\end{abstract}

\begin{IEEEkeywords}
Processor Architecture, Partitioned System, Mixed-criticality System, Real-time System, FPGA, Integrated Modular Avionics, Flight Computer
\end{IEEEkeywords}

%
\IEEEpeerreviewmaketitle


\section{Introduction}
%
%
%
%


\IEEEPARstart{M}{}\new{ixed-criticality implementation 
is gaining attention in safety-critical real-time cyber-physical systems. Driven by space, weight, power and cost constraints such embedded platforms can adapt to a mixed-criticality implementation where application software(s) with different criticality-level can share the same execution hardware with an isolation mechanism, in spatial as well as temporal domain, to restrict adverse interference between executing software.} 

Integrated Modular Avionics (IMA)\cite{DO297}, mixed-criticality system for airborne platforms, requirements allow application software of different criticality levels, i.e. design assurance level (DAL A-E) according to avionics  software guidelines DO-178B \cite{DO178B}, to share the same computational platform with established isolation in \textit{temporal} and \textit{spatial} domain; temporal isolation preserves execution timings and spatial isolation preserves the execution states of the applications belonging to different DAL levels. 

In conventional practice, such isolation is often achieved with a hypervisor, a \textit{software-based isolation} that creates virtual partitions on the hardware platform to accommodate applications with different criticality-levels in separate partitions. However, the hypervisor itself requires the same level of certification as the application software with the highest level of criticality running within the hypervisor. Additionally, functionalities rely upon the exposed features on the hardware, which restricts the selection of hardware platform. Besides, such certifiable/ certified products are expensive and often use proprietary tool-chain that needs specific skills. Larger platforms like civil airliner can afford such expensive development process and software overhead, however, the same is not optimal for smaller airborne platforms like \textit{unmanned aerial systems (UAS)}. 
\new{Furthermore, most of the real-time embedded systems, if not all, are designed to serve specific design goals and to operate for years with rare or no modifications for the entire lifespan. Unlike customization extensive general-purpose computing, such operational conditions give the opportunity of {improvement} by designing more hardware specific system design and cut software overhead. 

A hardware-based solution can potentially remove the software-overhead by providing isolation at the hardware level, allowing applications of different criticality-level to execute within \textit{partitions} without any software support for \textit{virtualization}}. One primary requirement for such a hardware platform is to provide analyzable timing behavior for the executing software. The timing constraints for a hard-real-time application is crucial as software timing behavior is a part of correctness, and a deterministic \textit{worst-case-execution-time (WCET)} must be established to guarantee timely completion of execution. 
Resource sharing between applications increases interference and potentially affects timing behavior unless isolation is established in the temporal domain\cite{multicore}.  


\subsection{Background}
\rev{The studies \cite{43_5}\cite{EASA}, reveal that a software defined processor synthesized on FPGA logical threads can potentially meet the hardware component requirements for airborne systems \cite{DO254}. Hardware-based isolation can be achieved by deploying applications of different criticality-level on separate hardware components: either separate processors, separate cores in multi-core, or separate threads on a multi-threaded processor.

Separate cores connected over an inter-core communication channel such as a \textit{network-on-chip (NoC)} can offer a \textit{single-core-equivalent-multicore} platform where each core can be dedicated to a certain criticality level. Such \textit{asymmetric multiprocessing} architecture can offer excellent solutions for parallel computation as well as scalability. Although, the scope of parallelization in control application is limited as the control flow: \textit{sensing-computing-actuating} is a sequential process. Schoeberl et al in \cite{SCHOEBERL2015449}, demonstrated a single core equivalent multi-core architecture with multiple cores interlinked with a time-analyzable  NoC. The architecture is demonstrated for IMA implementation in \cite{7445422}, where individual applications are implemented on separate processors to prevent interference. In \cite{66cfcda7e249495d974555dc889b742e}, an asymmetric multiprocessing architecture is demonstrated for improvement in overall system reliability. However, such architecture can impose significant hardware overhead that may limit its implementation in small airborne platforms. } 

In an alternative approach, hardware level isolation can be achieved by leveraging hardware threads in multithreaded processors. Hardware-level spatial isolation is maintained between tasks deployed on separate hardware threads within the processor and temporal isolation is implemented with a scheduler. A coarse-grained multithreaded processor targeted to improve overall throughput by efficiently utilizing the resources by switching thread when a thread is stalled for dependency e.g. \textit{cache miss}. However, WCET on such architecture is hard or impossible to determine due to uncertainties from dependencies such as \textit{memory}. 

\new{On the other hand, in fine-grained multithreading, each hardware thread is given access for a single clock cycle and instruction from each thread interleaves the pipeline at every clock cycle \cite{6925994}. 
The fine-grained multithreading can potentially facilitate excellent resource utilization with time analyzable execution, however, one major drawback in this architecture is demanding memory access requirements, to accommodate interleaving instructions from different threads in every processor clock cycle, which can only be met with an expensive scratchpad memory or separate memory devices dedicated to each thread.}

Although, a fine-grained multithreading offers a better concurrency, which is beneficial for inter-systems interactions such as IO-handling, the  WCET of executing applications scales up by a factor $n$ as compared to WCET on a single threaded processor where $n$ is the number of threads.
The timing requirements is crucial in aerospace applications, like deadline in flight control applications are in the order of milliseconds \cite{multicore}, which is not hard to achieve by the modern processor with clock frequencies in nanoseconds; however, the length of input-output execution path, which is critical for WCET analysis, may vary depending on platform requirements, and a longer execution path will result in further higher WCET on a fine-grained multithreaded architecture. At the task level, the increased WCET may not be a matter of concern due to the extended deadline, but this becomes severe when the same concept is extended to partitions. 

Furthermore, depending on the functionality, the applications may have different execution periodicity requirements and the partitions, to facilitate the required execution periodicity to the application running with it, requires different cycle frequency in a cyclic execution. Scheduling to meet such requirements is hard to achieve or even impossible with a fine-grained multithreading architecture. 


\subsection{Contribution}

\new{In this work, we demonstrate a custom soft-processor core, \AE r\o \space (named after a small Danish island in the Baltic Sea), with inbuilt virtualization mechanism at the hardware-level, that can be interfaced with COTS memory devices such as a SRAM or SDRAM, to conform mixed-criticality requirements without any software-based protection mechanism such as a hypervisor or an operating system (OS). 

\vspace{3mm}
The specific contributions of the work include - \vspace{2mm}
\begin{itemize}
    \item {A single issue multithreaded real-time processor architecture featuring controllable replicated data-paths for inbuilt spatial isolation between partitions, with hardware-defined partition specific memory-interactions.} \vspace{2mm}
    \item {Custom co-processor support for time-analyzable non-preemptive partition scheduling and switching mechanism with minimal and constant switching overhead, offering cycle-accurate WCET analyzability for an individual application executing in different partitions.}\vspace{2mm}
    \item {A custom 16-bit register-register ISA for efficient instruction-memory utilization for resource constrained platforms.}\vspace{2mm}
\end{itemize}

Safety-critical application software can run on a certified host software such as an RTOS or directly on the hardware (bare-metal) in one partition, and another low-criticality payload application software can run on a generic OS or directly on the hardware in a separate partition without any interference. All custom features are made accessible with standard C99 and existing and available tool-chain. 
In this work, we have demonstrated the processor with \textit{three} partitions for accommodating applications with three different levels of criticality that can be classified as {DAL A, B} and {C} levels. The hardware is defined in {Verilog} HDL and synthesized on an Intel Cyclone V FPGA for demonstration. }




%

\section{Design}
   
The \AE r\o \space processor is a 32-bit custom ISA, RISC style soft-processor with a four-stage pipeline: instruction fetch (F), decode (D), execute (E), memory-access (M).
All the components are designed with non-blocking sequential logic for easy timing analysis and to support fast clocking. The ISA, processor and co-processor architectures and the memory hierarchy are discussed in the following subsections:  
 

\subsection{Instruction Set Architecture}
In this work, we introduce a 16-bit custom \textit{register-register} instruction set architecture (ISA) for ease of analysis and debugging, however, the ISA has no contribution to the hardware-defined partition mechanism. 

\new{Driven by the traceability requirement in airborne software development guidelines, the constants and parameters in avionics software source-code are defined and initialized separately from their usage \cite{DO178B, DO178C}. This allows us to develop a fully functional ISA without \textit{immediate} instructions. Further, the ALU operations are limited to internal registers only. These two considerations give us the freedom to use a 16-bit ISA instead of a conventional 32-bit ISA, allowing the use of smaller and cheaper COTS memory devices as instruction-memory.} 

\begin{figure}[h]
\begin{center}
\includegraphics[scale= 0.5]{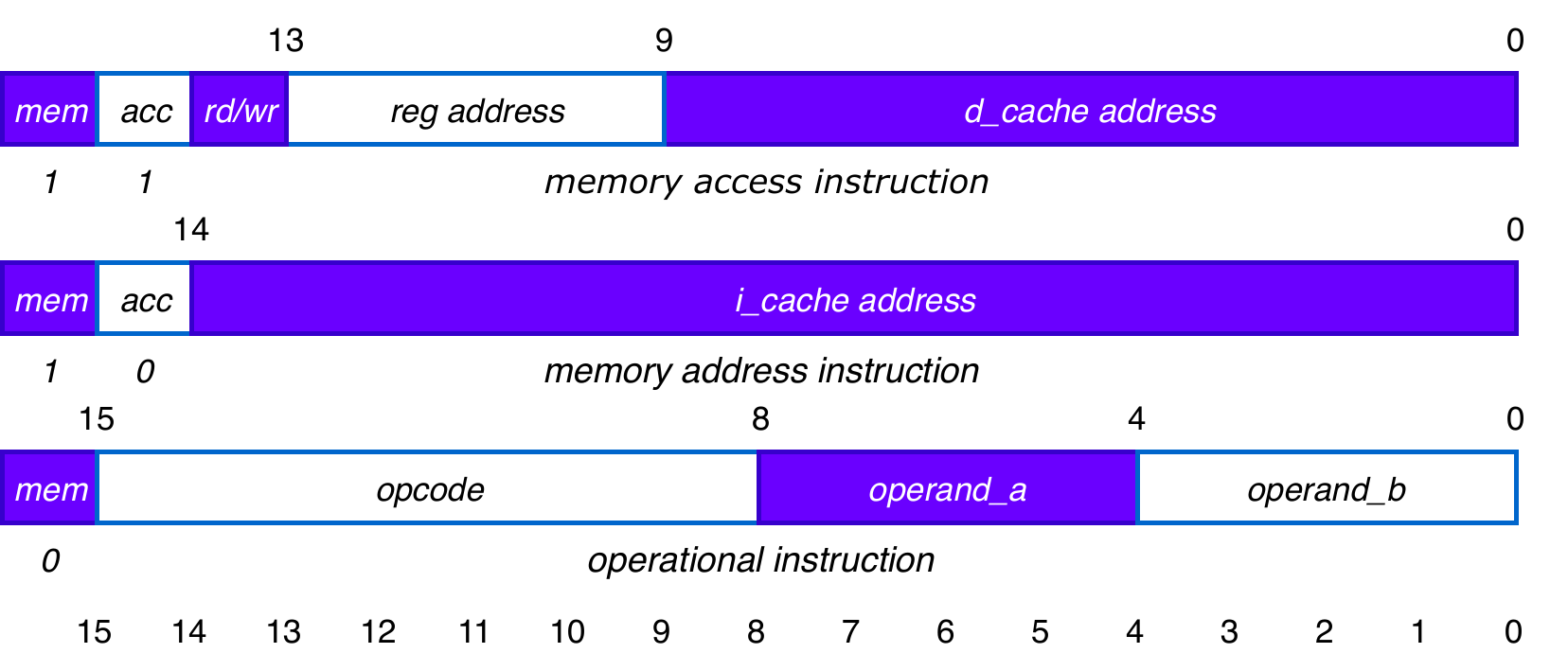}
\caption{Three types of 16-bit instructions:\textit{ memory-access instruction, memory-address instruction} and \textit{operational instruction}.}
\label{fig:instructions}
\end{center}
\end{figure}

The custom ISA uses three types of instructions: \textit{memory-access instructions},  \textit{memory-address instructions} and \textit{operational instructions} as shown in Figure \ref{fig:instructions}. 

\subsubsection{Memory-access instructions}
\new{ are the only instructions to access data memory to perform load and store operations to move content between data cache and internal registers. In a \textit{memory-access instruction}, the highest (MSB) two bits are hard-coded to $1$, which is unique to this instruction. The 14th bit in \textit{memory-access instructions} represent the direction of the data flow, i.e. load or store operation, where $1$ represents a store operation. The next 4 bits represent the internal register address and remaining 9 bits hold the address of the memory location in data cache, to perform a load or a store operation.}  

\subsubsection{Memory-address instructions}
 are used to point jump location in the instruction memory for branch and fork operations. In a \textit{memory-address instruction} the highest (MSB) two bits are hard-coded to $1$ and $0$, followed by a 14-bit instruction cache address.

\subsubsection{Operational instructions} are used to perform arithmetic and logical operations. The MSB in an \textit{operational instructions} is hard-coded to $0$, followed by a 7-bit \textit{opcode}, followed by two internal operand register identifiers. The \textit{operational instructions} only operate on internal registers,  which allow us to remove destination address from the \textit{operational instructions} and define a convention that first operand register is always the destination register. The ISA can accommodate up to 128 unique opcodes; some useful \textit{opcodes} are listed in Table \ref{Tab:opcode}.

However, the custom 16-bit instruction-set has a drawback as the branch and fork operations take two instructions, where the jump address is set with a memory-address-instruction and the branch (or unconditional jump) condition is evaluated with a successive operational-instruction.  

\renewcommand{\arraystretch}{1.1}
\begin{table}[ht]
\begin{center}
\caption{Selected Operational Instructions (OPcodes)}
\label{Tab:opcode}
\begin{tabular}{ccl}
\hline
\hline

\multicolumn{1}{c|}{\textit{Encoding}} & \multicolumn{1}{c|}{\textit{Mnemonic}} & \multicolumn{1}{c}{\textit{Operation}}    

\\ \hline

\multicolumn{1}{c|} {\fontfamily{qcr}\selectfont 0x11}   &\multicolumn{1}{c|} {add}                  & {\fontfamily{qcr}\selectfont op\_a = op\_a + op\_b}\\
\multicolumn{1}{c|} {\fontfamily{qcr}\selectfont 0x12}   &\multicolumn{1}{c|} {sub}                  & {\fontfamily{qcr}\selectfont op\_a = op\_a - op\_b} \\
\multicolumn{1}{c|} {\fontfamily{qcr}\selectfont 0x13}   &\multicolumn{1}{c|} {mul}                  & {\fontfamily{qcr}\selectfont op\_a = op\_a * op\_b} \\
\multicolumn{1}{c|} {\fontfamily{qcr}\selectfont 0x31}   &\multicolumn{1}{c|} {xor}                  & {\fontfamily{qcr}\selectfont op\_a = op\_a \^{} op\_b} \\
\multicolumn{1}{c|} {\fontfamily{qcr}\selectfont 0x32}   &\multicolumn{1}{c|} {and}                  & {\fontfamily{qcr}\selectfont op\_a = op\_a \& op\_b} \\
\multicolumn{1}{c|} {\fontfamily{qcr}\selectfont 0x33}   &\multicolumn{1}{c|} {or}                   & {\fontfamily{qcr}\selectfont op\_a = op\_a | op\_b} \\
\multicolumn{1}{c|} {\fontfamily{qcr}\selectfont 0x34}   &\multicolumn{1}{c|} {shr}                  & {\fontfamily{qcr}\selectfont op\_a = op\_a >> op\_b}  \\
\multicolumn{1}{c|} {\fontfamily{qcr}\selectfont 0x35}   &\multicolumn{1}{c|} {shl}                  & {\fontfamily{qcr}\selectfont op\_a = op\_a << op\_b} \\
\multicolumn{1}{c|} {\fontfamily{qcr}\selectfont 0x21}   &\multicolumn{1}{c|} {jle}                  & {\fontfamily{qcr}\selectfont j\_en = (op\_a \textless{}= op\_b) ? 1:0} \\
\multicolumn{1}{c|} {\fontfamily{qcr}\selectfont 0x22}   &\multicolumn{1}{c|} {jge}                  & {\fontfamily{qcr}\selectfont j\_en = (op\_a \textgreater{}= op\_b) ? 1:0} \\
\multicolumn{1}{c|} {\fontfamily{qcr}\selectfont 0x23}   &\multicolumn{1}{c|} {jl}                   & {\fontfamily{qcr}\selectfont j\_en = (op\_a \textless{}  op\_b) ? 1:0}\\
\multicolumn{1}{c|} {\fontfamily{qcr}\selectfont 0x24}   &\multicolumn{1}{c|} {jg}                   & {\fontfamily{qcr}\selectfont j\_en = (op\_a \textgreater{}  op\_b) ? 1:0} \\
\multicolumn{1}{c|} {\fontfamily{qcr}\selectfont 0x25}   &\multicolumn{1}{c|} {je}                   & {\fontfamily{qcr}\selectfont j\_en = (op\_a == op\_b) ? 1:0}\\
\multicolumn{1}{c|} {\fontfamily{qcr}\selectfont 0x26}   &\multicolumn{1}{c|} {jne}                  & {\fontfamily{qcr}\selectfont j\_en = (op\_a != op\_b) ? 1:0} \\
\multicolumn{1}{c|} {\fontfamily{qcr}\selectfont 0x27}   &\multicolumn{1}{c|} {juc}                  & {\fontfamily{qcr}\selectfont j\_en = 1}     \\
\hline              
\end{tabular}
\end{center}
\end{table}
\renewcommand{\arraystretch}{1}

\begin{figure*}[ht]
\begin{center}
\includegraphics[scale= 0.55]{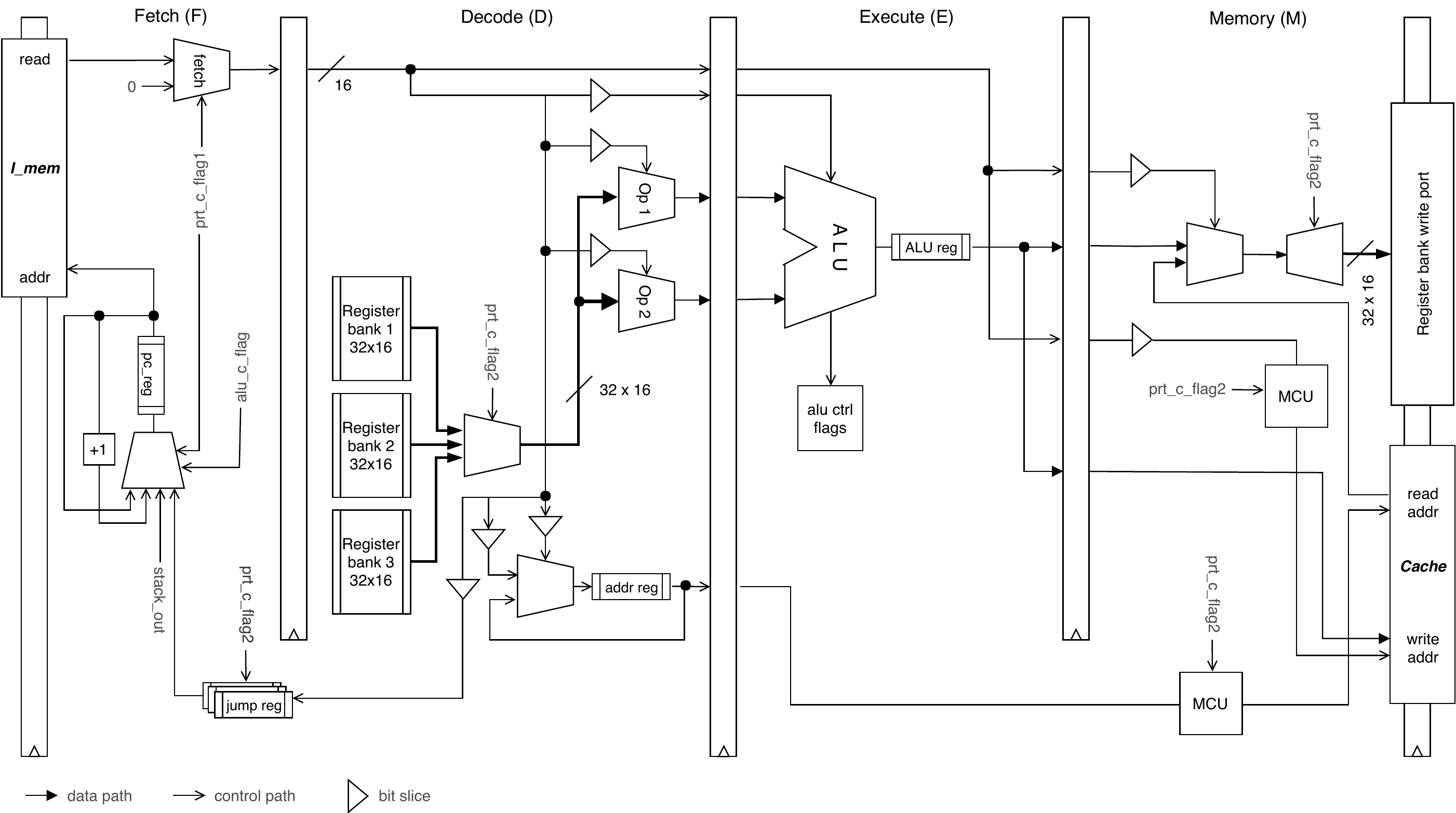}
\caption{A high-level diagram showing \AE r\o \space processor architecture and pipeline stages. Note that data paths and control paths are illustrated with different arrow pointers.}
\label{fig:CPU}
\end{center}
\end{figure*}

\subsection{\AE r\o \space Microarchitecture}
Microarchitecture of the \AE r\o \space processor is presented in Figure \ref{fig:CPU} that implements the ISA discussed earlier. The design goal of the processor core is to facilitate isolation between instructions flow from different partitions, and a cycle-accurate partition switching mechanism without losing time-analyzibility of the software running within each partition. Once access to the resources is given back to a partition which has executed earlier, should be able to resume execution from the exact same state for computational correctness; skipping an instruction or repeated execution of an instruction can potentially lead to erroneous computation. One way of saving the state of a partition is to copy the internal registers in the memory before switching to the next partition and copying back the contents from the memory to the registers when execution control is given back to the partition, as followed in several software-based isolation. 
To establish complete isolation between partition threads, resources are replicated to avail dedicated resources to the partitions. Each partition is allocated a separate register bank, stack address registers, a jump address register and a program counter (pc) register. The replicated resources are connected to the pipeline through multiplexers, where the resources are activated when the associated partition has execution access. 
The pipeline flow has no control in activating or deactivating the resources associated with a partition. In fact, the instruction-set does not carry any information or sense about the replicated resources or partitioning. 

\vspace{2mm}
\subsubsection{Fetch} The \textit{fetch stage} pushes a new instruction in the pipeline at every clock cycle. The fetch line is connected to the instruction cache and a hard-coded no-op instruction with a multiplexer that is controlled by partition switching mechanism. When the control line is active, no-op instructions are pushed into the pipeline, otherwise, instruction from the instruction cache is inserted. 
A program counter module keeps track of the instruction to be read in the next clock cycle by controlling the instruction memory address stored in a program counter register, {\fontfamily{qcr}\selectfont pc\_reg}. 
\vspace{2mm}
\subsubsection{Decode} In \textit{decode stage}, the 16-bit instruction is bit-sliced or decoded as per the ISA. \new{For operational-instruction, the opcode and two operand registers from the register bank is selected by setting the control line controlling the multiplexers {\fontfamily{qcr}\selectfont Op\_1} and {\fontfamily{qcr}\selectfont Op\_2} as shown in Figure \ref{fig:CPU}.} The instruction cannot select any specific register bank, and it only selects the operand registers from the register bank which is active. In case of a memory-address-instruction, the jump register is set to the branch address. For a memory-access load-instruction, the data cache address is set for reading, where, for a store access, the internal source register is selected as first operand with associated opcode. 
\vspace{2mm}
\subsubsection{Execute} In \textit{execute stage}, the ALU operations are performed on the operands selected in the decode stage and the result is stored in the {\fontfamily{qcr}\selectfont alu\_reg} for arithmetic operations or control flags ({\fontfamily{qcr}\selectfont alu\_ctrl\_flags}) are set for logical operations. Furthermore, the destination internal register (i.e. the first operand register) is selected for write back. All ALU operations are single cycle and \textit{atomic}; for multiplications, we have used DSP multiplier blocks for single cycle operation and hardware division is not featured in this work.
\new{A memory-address-instruction has no operation in the execution stage whereas, for a memory-access-instruction (i.e. \textit{load} and \textit{store}), the data cache address to be written or read is set to the destination memory address carried in the memory-access-instruction in the execute stage. As the memory-access-instructions do not contribute to ALU operations, using the execution stage for source and destination addressing do not affect the pipeline flow.

\vspace{2mm}
\subsubsection{{Write-back/ Memory-access}} The register-register ISA permits to implement the memory-access and the write-back operations in a single stage. Note, the data-in port of the internal register banks is connected to the ALU output as well as to the data cache output, and the ALU output is connected to data-cache \textit{datain} port as well as the data-load port of the internal registers.
The \textit{write-back} stage implements either of the two operations depending on the type of instruction: for an operational-instruction, the computational results are written back to the internal register from the {\fontfamily{qcr}\selectfont alu\_reg}, and for memory-access-instruction, the data is either copied to internal register from data-cache (load operation) or data is written to a memory location in the data-cache from internal register (store operation). All the writing procedures are non-blocking, single cycle and atomic. }

For simplicity, there is no port-forwarding mechanism in the processor, and potential data hazard is addressed with inserting \textit{bubbles or no-op} in the pipeline.




\vspace{2mm}
\subsubsection{{Branching}}
Due to limited space in a 16-bit of instruction, the branching operation requires two subsequent instructions. The first instruction is a memory-address instruction that sets the {\fontfamily{qcr}\selectfont jump\_reg} to the branch address. The following instruction is an operational-instruction that evaluates the branching condition.  The branch prediction is implemented as \textit{predict-not-taken} such that, if the branching condition is met, a jump enable signal, {\fontfamily{qcr}\selectfont j\_en}, is held high for a clock cycle that triggers the copying of {\fontfamily{qcr}\selectfont jump\_reg} to {\fontfamily{qcr}\selectfont pc\_reg}.

\vspace{2mm}
\subsubsection{{Subroutine call}} 
Similar to \textit{branching}, subroutine call requires two instructions. 
A hardware subroutine return register is featured in the processor core, which is set to the return address each time the subroutine call is performed. Also, the return address, i.e. the \textit{call instruction address} + 1 is stored in a stack and the stack pointer is increased by unity. In this work, we have used two separate stack pointers; {\fontfamily{qcr}\selectfont stack\_read\_pointer} and {\fontfamily{qcr}\selectfont stack\_write\_pointer} where the former points to the return address of the current subroutine call and latter points to the address where next return address will be written unless the stack is freed. A dedicated stack for each partition is used to keep track of the subroutine calls and at any instance, and the return address is set to the stack output. Once \textit{return} instruction is executed, {\fontfamily{qcr}\selectfont PC\_reg} is set to the stack output and {\fontfamily{qcr}\selectfont stack\_read\_pointer} and {\fontfamily{qcr}\selectfont stack\_write\_pointer} are decremented by unity. 

The subroutine call is performed by two instructions, where the first instruction sets the subroutine address in the instruction memory, followed by an operational-instruction that triggers the {\fontfamily{qcr}\selectfont call\_en} flag, which is one of the CPU control flags.

\vspace{2mm}
\subsubsection{{Pipeline flushing}}
As the WCET is the effective computation time in real-time systems, there is no branch prediction algorithm implemented, and branching is done at the end of the execution cycle without any prior prediction. Hence, whenever a branch is taken, the instructions in the {\fontfamily{qcr}\selectfont fetch\_reg} and {\fontfamily{qcr}\selectfont decode\_reg} belong to the flow from the other branch (the default case) and need to be flushed. The same is applicable for subroutine calls as well. Flushing is only required in the fetch and decode stage and flow in the pipeline from decode to execute stage can be prevented with a \textit{stall} for a clock cycle.  
Furthermore, the {\fontfamily{qcr}\selectfont fetch\_reg} needs to be flushed at the next clock-cycle, as fresh instruction comes out of the instruction cache in the subsequent clock cycle after setting the address register to the desired value. 
In \AE r\o , \textit{flushing} is auto triggered when {\fontfamily{qcr}\selectfont j\_en} or {\fontfamily{qcr}\selectfont call\_en} are set to active.  
\vspace{2mm}

The processor facilitates two especial control lines; {\fontfamily{qcr}\selectfont ptr\_c\_flag1} and {\fontfamily{qcr}\selectfont ptr\_c\_flag2} which are used for establishing temporal and spatial isolation respectively. The former control line is connected to the program counter and instruction fetching mechanism to control the inflow in the pipeline, and the latter is connected to the register banks interfaces and memory control unit (discussed later) to maintain isolation between data flow from separate 
partitions. The {\fontfamily{qcr}\selectfont ptr\_c\_flag2} is always set to the active partition index which controls the active data paths for each partition.


Furthermore, the processor facilitates an exposed hardware timer that is accurate to the processor clock cycle and an active partition id register, that is set to the executing partition id by the processor. The $32\times2$ bit clock resets after $2^{64}$ clock cycles, i.e. in the order of decades and virtually cannot reset within runtime. 
Both the hardware are memory mapped, and can be read by reading the memory-mapped locations without any special set of instructions. 

\subsection{{Partition Switching}}
The partition switching mechanism is triggered by a co-processor, Switching-Control-Unit (SwCU). The SwCU is a cycle accurate time-triggered arbitration mechanism that implements scheduling set by the system designer for each partition. The SwCU can accommodate different execution period and execution time for each partition, however, to facilitate time analyzable behavior the following assumptions are made - 
\begin{itemize}
    \item \textit{All partitions are periodic}. 
    \item \textit{All switching events are time-triggered}.
    \item \textit{All partitions have uniform priority}.
\end{itemize}

Note, the uniform priority should not be misinterpreted as uniform criticality; uniform priority suggests that any two partitions shall never compete for execution access.

\begin{figure}[h]
\begin{center}
\includegraphics[scale= 0.45]{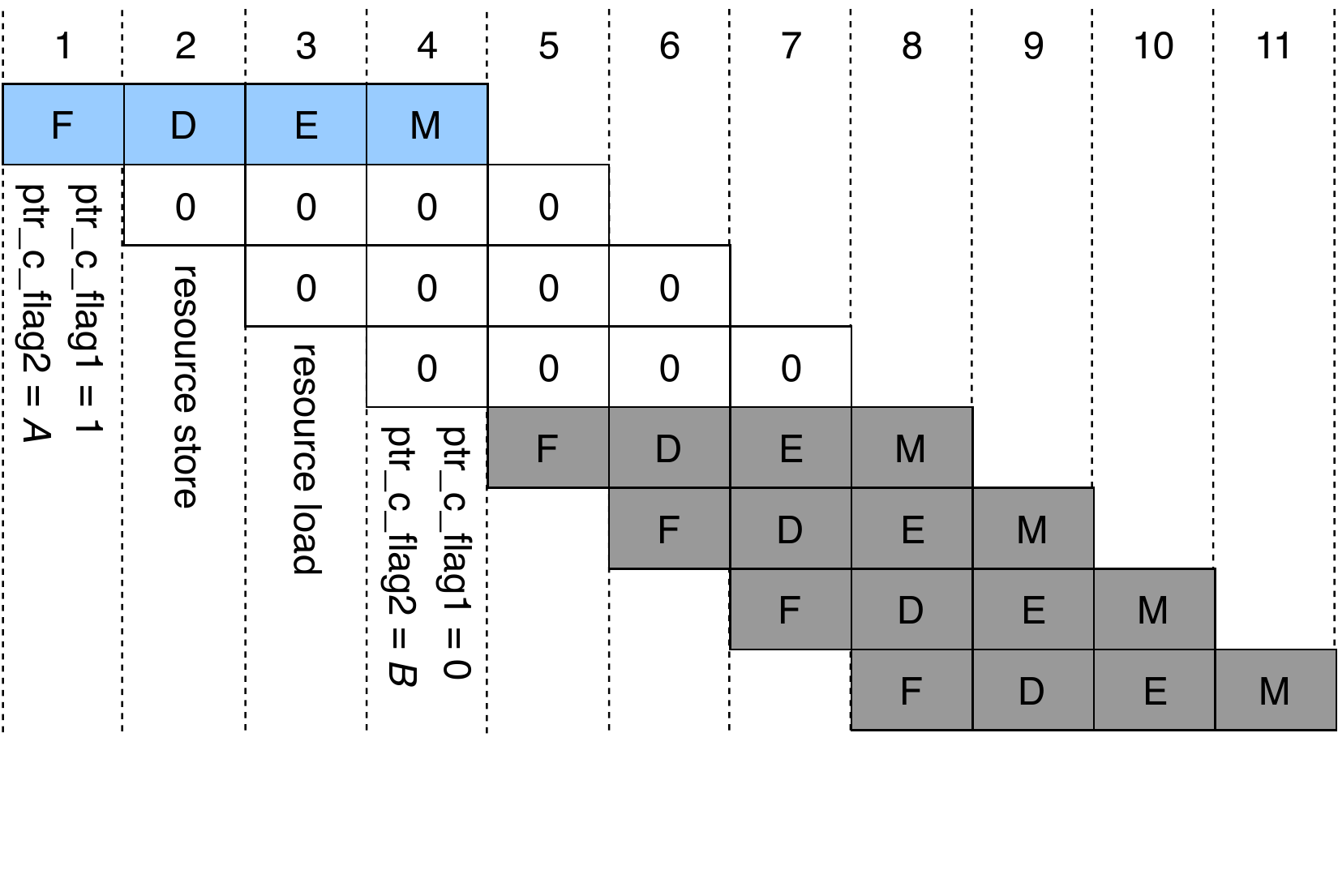}
\caption{Showing the pipeline flow during a partition-switching operation (\textit{A} to \textit{B}). Note that the control signals are effective in the subsequent clock cycle due to non-blocking sequential logic.}
\label{Fig:flow}
\end{center}
\end{figure} 

Instead of a look-up table the SwCU operates on configurable \textit{period} and \textit{execution time} for each partition which needs to be defined by the system designer. 
There could be dependencies between the partitions in terms of order of execution. Such dependency is accommodated by setting a starting sequence during initialization of the system. 
The SwCU features separate period clocks for each partition and a shared execution clock. At power-up or reset condition, the period clocks are initialized as per the starting sequence set by the system designer. Each period clock is decremented by unity in each clock cycle whereas the execution clock is incremented by unity. 
Once a partition is given execution access, the associated period clock for the partition is reset to its period cycle time and the execution clock is set to zero; once the execution clock reaches the pre-configured execution time for the partition, an expiry\_flag is set to denote the expiry of the execution time.  

Similarly, when a period clock of a partition reaches a preset value i.e. partition switching time (a constant value) the partition switching procedure starts; the program counter is stalled by activating {\fontfamily{qcr}\selectfont ptr\_c\_flag1} control line and the current program counter content is stored in the pc\_register associated to the active partition. While {\fontfamily{qcr}\selectfont ptr\_c\_flag1} is active, the fetch stage inserts \textit{no-op} instructions in the pipeline. The existing instructions in the pipeline are uninterrupted and the pipeline flow is maintained (no stalling in the pipeline). Once the period clock reaches to value $1$, the {\fontfamily{qcr}\selectfont pc\_reg} is set to the new value from the next partition. In the following clock cycle {\fontfamily{qcr}\selectfont ptr\_c\_flag2} is set to the next partition index and {\fontfamily{qcr}\selectfont ptr\_c\_flag1} is deactivated as shown in Figure \ref{Fig:flow}.


\subsection{{Memory Hierarchy and I/O Interfacing}}
The memory management must accommodate isolation and protection mechanism to the associated memory regions to each partition to prevent erroneous access by other partitions. Shared memory can be used to share data between partitions when some access control mechanism is implemented to prevent data-corruption. Such corruption may occur if a memory location written by a producer gets over-written by an application in another partition before the intended consumer application reads it. In this work, we have used three types of memories: instruction cache, data cache and stack memory. All memory operations are single cycle.
\vspace{2mm}
\subsubsection*{\textbf{Memory-Control-Unit}}
To establish isolation in the memory regions without any software support a memory-control-unit (MCU) is defined on the hardware to segment memory region. The MCU is directly driven by the hardware-defined partition switching mechanism ({\fontfamily{qcr}\selectfont ptr\_c\_flag2}) and controls the two most-significant-bits of the memory address accordingly to the active partition. If the physical address length of a memory unit is $n$-bits, the $[n-1:n-2]$ bits are controlled by the MCU and $[n-3:0]$ bits are controlled by the CPU. Software running within a partition has no notion of the memory segmentation. 
\vspace{2mm}
\subsubsection*{\textbf{Data and Instruction Cache}}
The instruction and data caches are separate and independent. \rev{For cache memories the following assumptions are made - 
\begin{itemize}
    \item \textit{All instructions can be accommodated on the instruction cache}. 
    \item \textit{All data can be accommodated on the data cache}.
    \item \textit{There is no cache miss}.
\end{itemize}
Such assumptions are valid for real-time safety-critical systems, where all the constants and variables are pre-defined and pre-initialized. Cache miss detection is out of the scope of this work. }

The 16-bit instruction cache is configurable and can be extended to off-chip implementation with external SRAM chip. The CPU has no physical connection to the instruction cache {\fontfamily{qcr}\selectfont write\_enable} line and cannot modify any memory location in the instruction cache and a single memory device can be used for all the partitions. The 14-bit address space can hold 32K bytes of instruction per partition (with MCU to extend the address space to 16-bit), which we have considered sufficient for this work and haven't considered any extension for instruction memory.

The 32-bit dual port data cache is defined on on-chip block memory. 
All the partitions have write-access to the data cache and some protection mechanism is essential for isolation.
The data cache is divided into two regions; protected and shared. 
The protected region is further segmented into dedicated regions for each partition.
The dedicated regions can only be accessed by the partition it is allocated to and enforced by the MCU. 
The dedicated memory region is also used as a data stack and to store static data. 
The stack base is located at the low end that grows depending upon the application needs. Note, that dedicated memory region is controlled by the MCU and a software within a partition needs not to consider address overlapping and can use same address; for e.g. the stack base address for software in all the partitions are the same, but the MCU differentiates the physical location in the memory. 
All the partitions have access to the shared memory region and use it to share data. As only one partition is active at any given time, no simultaneous read-write protection is needed.
During power-up or system reset, the static data is copied to the data cache. 
The address stack is a separate single port 16-bit memory, defined on on-chip block memory. The address stack is only used to handle subroutine calls. 

\new{\subsubsection*{\textbf{I/O Interfacing}}
There is no architecture specific requirement for external \textit{input-output (I/O)} interfacing, especially when memory-mapped interfaces are used with sampling-buffers, which is the conventional practice in aerospace systems. Partition specific I/O ports can be memory-mapped in the partition specific restricted memory space to prevent access from other partitions, where, shared I/O ports between multiple partitions can be memory-mapped in the shared memory region. 

Streaming buffers, such as FIFO based solutions cannot be directly shared between partitions as it violates the concept of spatial isolation where the FIFO read pointer can be altered by a dysfunctioning application within any partition which may or may not supposed to do so, and such interfaces shall be mapped to a partition specific restricted memory region.

In practice, the processor is not directly interfaced with external I/O systems in safety-critical systems and additional intermediate co-processors are used for I/O interfacing; for aerospace systems, additional hardware units such as \textit{remote-data-concentrator (RDC)} or \textit{avionics-data-concentrator (ADC)} are used that perform checks and processing of I/O data before it is ready for computation. }


\section{System Development}
Software language and development tools may require certification depending upon its functionality and role. Moreover, a custom tool chain needs additional training and documentation for airborne software consideration. Considering these facts, this work is focused on utilizing standard language, compiler and tool chain for software development. However, an ISA specific compiler is not available for this platform, instead a standard \textit{x86} LLVM Clang Compiler is used to generate assembly code and the generated code is modified with a custom assembler to support the ISA before generating binary executables.

For experimentation, we have considered standard C99 without any modifications as the default language and open source \textit{Code::Blocks} 17.12 as the development software. The C code is compiled using \textit{x86} LLVM Clang Compiler with 1999 ISO C language standard to generate assembly.
\vspace{2mm}
\subsubsection*{Assembler}
The \AE r\o \space processor architecture is based on a custom ISA and an ISA specific assembler is needed to process the assembly code generated by the compiler. A custom assembler is written in Python to process the assembly code and generate the executable binary. 

There are certain differences between \textit{x86} and the proposed ISA. 
Firstly, the custom ISA does not support ALU operation directly on data cache out, however, this restriction is not followed in compiler generated assembly and to be addressed by the custom assembler. This issue is handled in the assembler by inserting an intermediate instruction to move the content from cache to an internal register {\fontfamily{qcr}\selectfont emt}, and perform the ALU operation with initial operand register and {\fontfamily{qcr}\selectfont emt}, as shown below:
\vspace{2mm}
\\{\fontfamily{qcr}\selectfont add eax, dword ptr [c]} (\textit{original instruction})
\\{\fontfamily{qcr}\selectfont mov emt, dword ptr [c]} (\textit{inserted instruction})
\\{\fontfamily{qcr}\selectfont add eax, emt}                 (\textit{modified instruction})
\vspace{2mm}

Note that {\fontfamily{qcr}\selectfont emt} is a custom register specific to our architecture and never used in the generated assembly; {\fontfamily{qcr}\selectfont emt} register is always available for the assembler.

Secondly, the branch and call instructions are implemented differently in this ISA with two instructions which must be addressed by the assembler. In this ISA, the jump address is set with the former and the jump conditions (or unconditional jump) is evaluated with the latter instruction and the assembly code is modified to accommodate this as shown below (note, \textit{cmp} instruction is removed and a new instruction \textit{jad} is added, which implies jump address.): 
\vspace{2mm}
\\{\fontfamily{qcr}\selectfont cmp	eax, dword ptr [b] }(\textit{original instruction })\\ 
{\fontfamily{qcr}\selectfont jle	.LBB3\_2} (\textit{original instruction})
\\{\fontfamily{qcr}\selectfont jad .LBB3\_2 }(\textit{modified instruction})
\\{\fontfamily{qcr}\selectfont jle	eax, dword ptr [b]} (\textit{modified instruction})
\vspace{2mm}

Thirdly, the possible data hazard is also handled in the assembler for easy determinism and timing calculations. Although there are matured and sophisticated techniques like path forwarding to avoid data hazard, we have mitigated this by inserting a no-op instruction where a data hazard is possible at a cost of wasting one clock cycle. 

Fourthly, numerical constants in immediate instructions of the x86 assembly are mapped to data memory and the assembler maps the memory location in the instruction instead of the value. The data is copied to the {\fontfamily{qcr}\selectfont emt} register from the memory location and computation is performed in the subsequent instruction as explained earlier. 


Lastly, a modified assembly file (to conserve the mnemonic for debugging), an executable binary file and a binary static data file is generated that is downloaded in the hardware. 
The codes for individual partition require to be assembled individually for correct placement in the memory.


\section{Evaluation}
The objective of the evaluation is to analyze the platform for computational correctness and timing-correctness of executing applications within partitions. Besides, the motivation of the work is driven by the requirements of mixed-criticality systems in an airborne platform, and the feasibility of implementation in an airborne system needs to be demonstrated.  
The evaluation of the \AE r\o \space processor core is done in two phases;
a test case is created to analyze the functionality and timing behavior of the platform and the system behavior and performance are analyzed against an avionics use case.

\subsection{Setup}
For experimentation and demonstration, the \AE r\o \space processor core is synthesized on an FPGA chip. The processor alone is not sufficient for operation and requires additional hardware support like IO interfacing, oscillator, memory, a reset meachanism etc. A system-on-chip is set up around the processor core with 16KB of instruction memory, a 50 MHz oscillator, 32-bit debug port with LEDs at [7:0], an UART IPCore (115200 bps), a reset button, and a monitoring hardware module that exposes internal switching mechanism through GPIO pins. 
The hardware, written in Verilog HDL, is synthesized with Intel Quartus Prime 18.1 to generate .\textit{sof} file. We have used \textit{Intel De10-Nano} development board with Intel Cyclone® V SE FPGA chip for experimentation. 
The board features a 50MHz inbuilt oscillator, which is used as the primary clock for the processor core and other hardware modules. 
However, the dual-core ARM Cortex-A9 processor on the SoC chip is not used for any purpose in this work. 
Once the board is configured with the .\textit{sof} file, the JTAG connection used to configure the board, is disconnected. The on board UART IPCore is used to upload the executable in the memory. 


For accurate timing analysis a digital-oscilloscope is connected to the exposed GPIO pins for cycle accurate analysis of the partition switching mechanism. The oscilloscope is connected to the 1-bit {\fontfamily{qcr}\selectfont ptr\_c\_flag1} and 2-bit {\fontfamily{qcr}\selectfont ptr\_c\_flag2} signal that drives the switching mechanism.

\subsection{Timing Analysis}
The timing behavior of individual partition should be evaluated for accuracy. The partition switching is driven by a hardware mechanism and each partition should pose cycle-accurate timing-behavior. Moreover, the partitioning can potentially affect the WCET of the individual applications running within a partition.
Say, the WCET of an application, $A$, when executing in a non-partitioned environment is $\tau_{A_0}$. If the same application is ported to this platform into a partition $n$ that has an execution time of $\tau_{p_n}$, then the effective WCET of the application, $\tau_{A_n}$, can be determined as: 
\begin{multline*}
    \tau_{A_n} = \bigg(\bigg\lceil{\frac{\tau_{A_0}}{\tau_{p_n}}}\bigg\rceil-1\bigg) \times E_p +  \bigg\{ \tau_{A_0} -  \bigg(\bigg\lceil{\frac{\tau_{A_0}}{\tau_{p_n}}}\bigg\rceil - 1\bigg) \times \tau_{p_n} \bigg\}
\end{multline*}
where, $E_p$ is periodicity of the partition $n$. 
\revtwo{
\subsubsection*{\textbf{Derivation}} Let's assume, a task within a partition completes execution in $n$th execution access of the associated partition.
The remaining execution time after $(n-1)$th access:
\begin{equation*}
\tau_{A_0} - (n-1) \times \tau_{p_n}    
\end{equation*}
and the time interval between the first access and $n$ th access is : $(n-1) \times E_p$.
Hence, effective time required to complete execution is- 
\begin{equation*}
(n-1) \times E_p + \tau_{A_0} - (n-1) \times \tau_{p_n}    
\end{equation*}
$n$ is the number of times the partition is given access, and an integer and can be represented as $\lceil{{\tau_{A_0}}/{\tau_{p_n}}}\rceil$. }


Note, when $\tau_{p_n}$ \textgreater{}= $\tau_{A_0}$, $\lceil{{\tau_{A_0}}/{\tau_{p_n}}}\rceil$ in the above equation becomes $1$, resulting in $\tau_{A_n} = {\tau_{A_0}}$ i.e. when the execution period of a partition is greater than the no-partition equivalent WCET of an application, the WECT of the application is unaffected.

WCET time analysis for individual application has been presented in several works \cite{KOSMIDIS2016287}\cite{Lv2010}\cite{Schoeberl2009}, and can be considered to determine $\tau_{A_0}$.

Let's consider the following code segment as the application $A$, that we will consider for periodic execution in separate partitions for timing analysis:\vspace{2mm}
{\fontfamily{qcr}\selectfont
\\int m = 1, i = 1; \\
int zro = 0, threshold = 10000;\\
uint8\_t *timer = (*uint8\_t) 0x019;\\
uint8\_t *p\_id = (*uint8\_t) 0x01A;\\
uint8\_t *uart = (*uint8\_t) 0x018;\\
int main()\{\\
\indent if(m == i)\{ \\
\indent \indent     uart = p\_id;\\
\indent \indent     uart = timer;\} \\
\indent m += i;\\
\indent if(m == threshold)\{\\
\indent \indent     uart = p\_id;\\
\indent \indent     uart = timer; \\
\indent \indent     m = zro;\} \\
\} \\
}

\begin{figure}[h]
\begin{center}
\includegraphics[scale= 0.35]{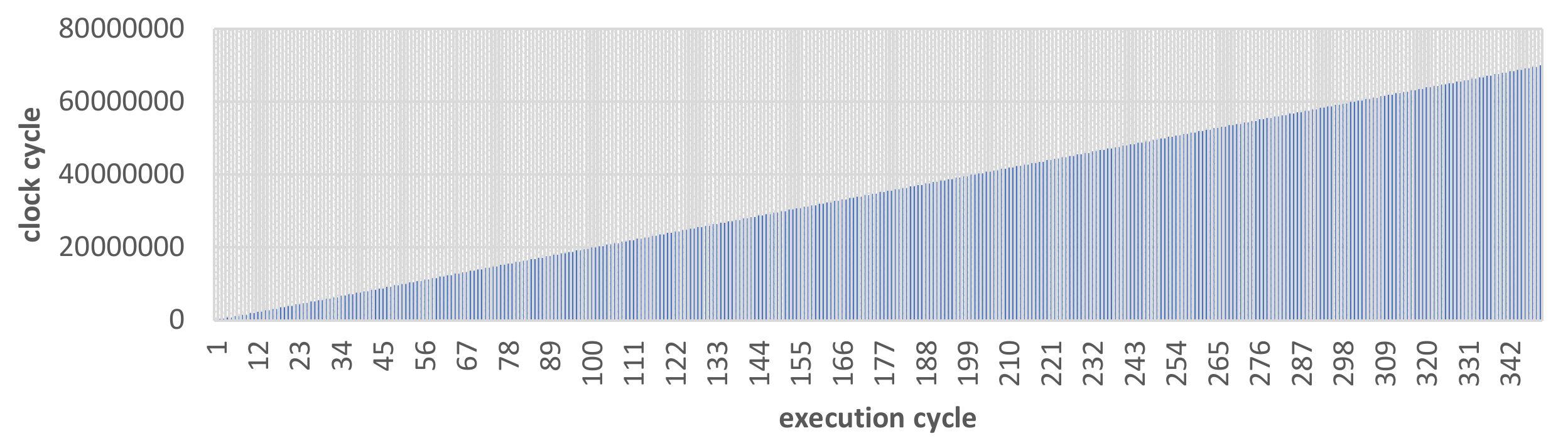}
\caption{Showing the uniform execution time for 350 subsequent execution cycle. Note, the bars represent timer reading for $m = i$ and $m = threshold$ as per the code fragment.}
\label{fig:exe}
\end{center}
\end{figure}

The \textit{timer, p\_id} and \textit{uart} are memory-mapped hardware modules mapped at memory address {\fontfamily{qcr}\selectfont 0x019, 0x01A} and {\fontfamily{qcr}\selectfont 0x018} respectively. The threshold value is determined to support the UART transmission rate. A fast transmission by the processor core will result in buffer overflow and packet loss. 
\new{When executed on a single partition with all other partitions deactivated, the execution time of computation cycles are identical and equal to $399963$ clock pulses or $7.99926$ ms i.e. $\tau_{A_0}$ as presented in Figure \ref{fig:exe}.}

\begin{figure}[h]
\begin{center}
\includegraphics[scale= 0.47]{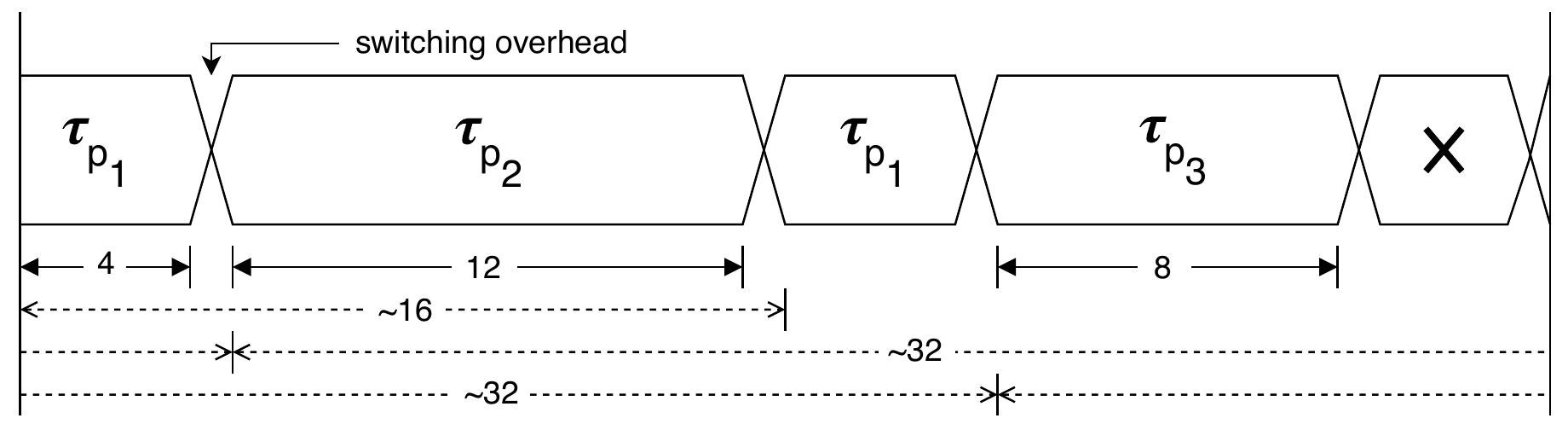}
\caption{Showing execution time and period allocated to three partition $p_1,p_2$ and $p_3$. The associated execution times are represented as $\tau_{p_1},\tau_{p_2}, \tau_{p_3}$. All the timings are in milliseconds. note, '$\sim$' is used to account the switching overhead of 10 clock cycles (i.e. $ 0.0002$ milliseconds). The 'X' represents the time slot when no partition is active.}
\label{fig:timing}
\end{center}
\end{figure}

\new{The same application is implemented in three separate partitions $p_1,p_2$ and $p_3$, with execution time of $\tau_{p_1} = 4 ms$, $\tau_{p_2} = 12 ms$ and $\tau_{p_3} = 8 ms$ to analyze the timing behavior. Note, that the execution time of partition $p_1$, $\tau_{p_1}$, is less than $\tau_{A_0}$. The partition execution schedule is presented in Figure \ref{fig:timing}. There could be applications specific requirements where no partition is active, to limit power consumption or meet exact periodicity, and to analyze the impact of such an implementation we have added a time slot (X) in the partition execution schedule where none of the three partitions are active.}

The Table \ref{Tab:ewcet} shows the scheduling implementation and clock cycles at which execution access given to each partition. $E_{start}$ is the clock cycle when access is given and $E_{end}$ is when access is taken back from a partition. Similarly, $\tau_{p_x}$ represents the execution access time in milliseconds of a partition $x$. Also, the outputs from the executing applications are presented in the table. Note that the outputs are the internal timer values sent by the executing application and not the time when the output is received at the monitoring end. 

\begin{figure*}[h!]
\includegraphics[scale= 0.56]{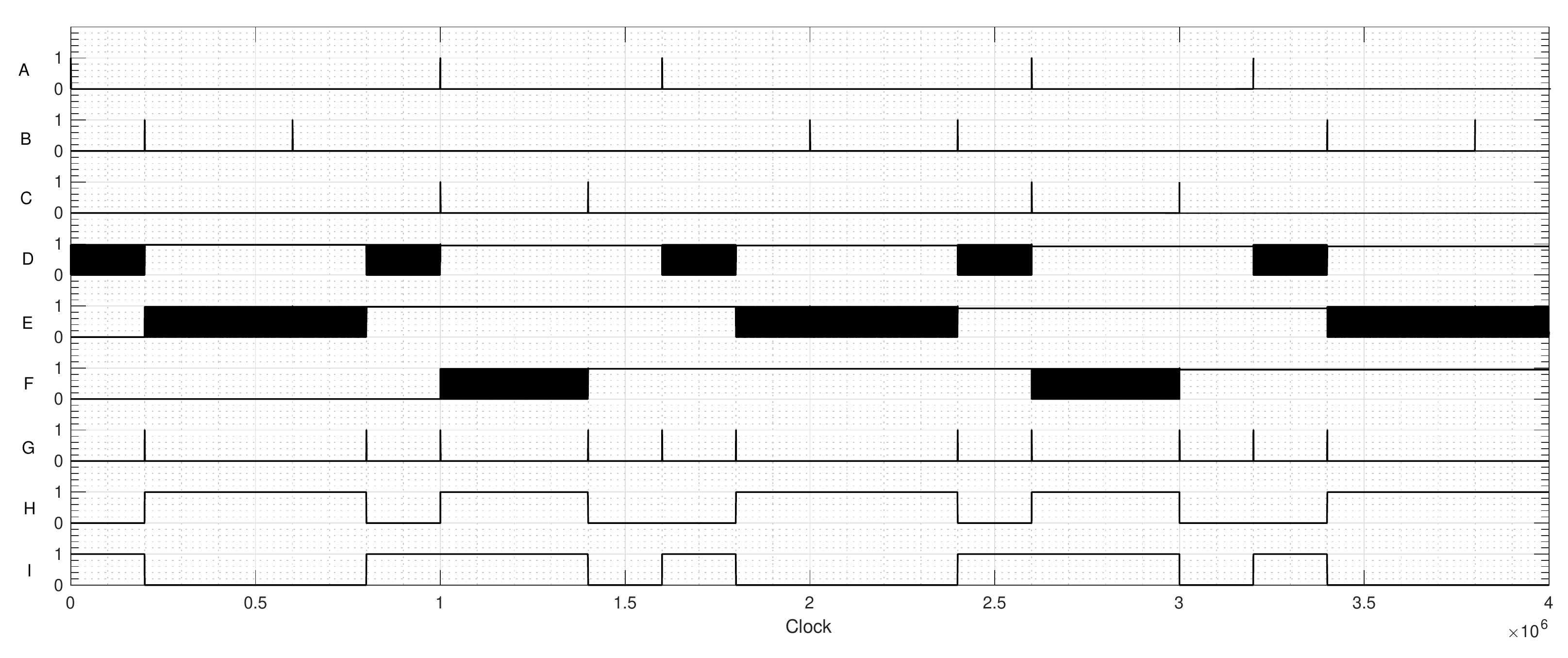}
\caption{Timing diagram showing cyclic execution within each partition and partition switching. At the top, (A, B, C) show the clock cycle when partition 1, 2 and 3 sends output to the UART IP respectively. (D, E, F) show the normalized values of program counters of partition 1, 2 and 3 respectively. Note that each program counter resumes from the exact same state when access is given back to the partition. (G) represents the $ptr\_c\_flag\_1$ flag and (H, I) represent the high-bit and low-bit of $ptr\_c\_flag\_2$ respectively.}
\label{fig:mcp}
\end{figure*}

\renewcommand{\arraystretch}{1.5}
\begin{table}[h!]
\begin{center}

\caption{Resource access time to partitions}
\label{Tab:ewcet}
\begin{tabular}{c|c|c|c|c}
\hline
\hline
Partition & \begin{tabular}[c]{@{}c@{}}$E_{start}$ \\ (cycle)\end{tabular}  & \begin{tabular}[c]{@{}c@{}}$E_{end}$ \\ (cycle)\end{tabular}  & \begin{tabular}[c]{@{}c@{}}$\tau_{p_x}$ \\ (ms)\end{tabular} & Outputs           \\ \hline
1         & 0            & 200000   & 4  & 16                           \\ \hline
2         & 200010       & 800010   & 12 & 200026,599969, 600027      \\ \hline
1         & 800020       & 1000020  & 4  &     999980         \\ \hline
3         & 1000030      & 1400030  & 8  & 1000046, 1399989           \\ \hline
x         & 1400040      & 1600030  & -- & --                           \\ \hline
1         & 1600030      & 1800030  & 4  & 1600068                          \\ \hline
2         & 1800040      & 2400040  & 12  & 2000020, 200078, 2400021               \\ \hline
\end{tabular}
\end{center}
\end{table}
\renewcommand{\arraystretch}{1}

\revtwo{
\renewcommand{\arraystretch}{1.5}
\begin{table}[h!]
\begin{center}

\caption{FPGA resource usage for partitioned vs non-partitioned implementation.}
\label{Tab:resource}
\begin{tabular}{c|c|c|c}
\hline
\hline
  & \begin{tabular}[c]{@{}c@{}}ALMs \\ \end{tabular}  & \begin{tabular}[c]{@{}c@{}} Combinational ALUTs  \\ \end{tabular}  & Registers           \\ \hline

Non-Partitioned         & 1607       & 1408    &  2277    \\ \hline
Partitioned (complete)      & 2653            & 2646     & 4493                           \\ \hline
Arbitrator        & 143            & 221    & 137                           \\ \hline

\end{tabular}
\end{center}
\end{table}
\renewcommand{\arraystretch}{1}
}

As shown in Figure \ref{fig:timing}, execution time for partition 2 and partition 3 is greater than $\tau_{A_0}$ ($\tau_{p_2} = 12 ms$ $\tau_{p_3} = 8 ms$), so the effective WCET for the application within partition 2 and 3 remains unchanged i.e. $\tau_{A_2} = \tau_{A_0}$ and $\tau_{A_3} = \tau_{A_0}$. However, as $\tau_{p_1} \textless{} \tau_{A_0}$, as per equation 1, the calculated $\tau_{A_1}$ is $19.99966 ms$ which is equal to the measured WCET from the output. A timing diagram of cyclic execution of the partitions and key control register states are shown in Figure \ref{fig:mcp}.

Although, it might seem unusual to have a partition execution time of a period less than the execution time of the applications running within it, due to the feasibility of scheduling applications with low priority and extensive resource demands can be scheduled on a partition with soft real-time boundations. 

\revtwo{\subsection{Comparison with Alternative Architectures}
Single-core-equivalent (SCE) multicore or many-core and fine-grained multithreaded (FGM) architectures can feature hardware-based isolation required for mixed-criticality systems and each has own merits and demerits as compared to \AE r\o \space. 

First, a single-core-equivalent multi-core architecture with same number of processing elements as the number of partitions used in \AE r\o \space can offer a platform for parallel execution. Three mutually independent tasks, $a, b$ and $c$, when ported to three processors in the SCE architecture, can execute simultaneously and the total execution time will be equal to the longest execution time among the three tasks. 
FGM and the \AE r\o \space platform do not support parallel computing. The total WCET of all tasks on \AE r\o \space is the sum of WCET of individual tasks and partition switching time, where, the same is 3x of the WCET of the longest task on an FGM architecture. The WCET of individual tasks on \AE r\o \space are same as the WCET on the SCE platform, in contrast, WCET of individual tasks are 3x on an FGM architecture. 

However, when the tasks are interdependent, for e.g. $a->b->c$, the scope of parallel execution is limited and the total WCET of the same tasks becomes the sum of individual execution time and inter-processor communication delay on a SCE architecture. In the same context, the total WCET on an FGM platform is the sum of 3x of the WCET of the individual tasks, where, irrespective of inter-dependency the total WCET remains same on the \AE r\o \space platform. 

Secondly, to assure bounded system response time, apart from WCET, periodicity of a system is critical. In mixed criticality systems, different tasks may require different execution frequencies and hardware platform plays a critical role in accommodating such requirements. In a SCE platform, a task within a processor can have a custom scheduling to meet its periodicity requirements irrespective of other tasks in other processors. Similarly, in \AE r\o \space, partitions can be scheduled to satisfy WCET and periodicity requirements of individual tasks. In contrast, an FGM architecture equally distributes the computational resources among the tasks in different threads, and tasks with equal WCET cannot have different periodicity. 

Lastly, each hardware architecture implements different features that results in different hardware overheads. The hardware overhead of a SCE architecture can be significantly higher than that of the partitioned architectures. The hardware requirement for each additional processing element linearly increases from what is required for a single un-partitioned processing element in addition to the hardware overhead for an inter-processor communication mechanism such as a NoC. A comparison of hardware resource requirements between a partition and a non-partition processing unit is presented in Table \ref{Tab:resource}.
The hardware overhead of a fine-grained multithreaded processor itself is very similar to the hardware overhead of the proposed architecture. 

Furthermore, for simultaneous operations in a SCE architecture, the requirement of separate instruction and data memory for each processor results in indirect hardware overhead when external memory devices (e.g. SRAM, SDRAM chips) are used. Similarly, in FGM architectures, to accommodate interleaving of instructions from different tasks in every clock cycle separate external memory devices are required that also results in additional interfacing (GPIO) requirements. 
In contrast, only one partition is active at any time in \AE r\o \space and external memory devices can be shared between partitions with isolation mechanism as proposed in this work while preserving spatial isolation. 

For execution of independent tasks, SCE architecture can outperform both fine-grained and \AE r\o , and offers the best solution for parallel execution at a cost of highest hardware overhead, where, an FGM architecture can feature a better concurrency and best suitable for event-triggered system, where an event/ interrupt belonging to any criticality level can be addressed in a short time. 
In contrast, the proposed architecture is novel for interdependent tasks with static scheduling. Although, such an architecture is inferior in terms of concurrent and parallel execution as compared to the other two, it can offer same, if not better, total WCET for multiple interdependent tasks as compared to other architectures with similar or less hardware overhead of an FGM architecture. A comparison between the proposed architecture vs alternative architectures that feature hardware-based isolation is presented in Table \ref{Tab:comp}.  }
\renewcommand{\arraystretch}{1.5}
\begin{table*}[]
\begin{center}
\caption{Comparison between the proposed architecture vs alternative architectures with hardware-based isolation. 
Note, $\tau$ is WCET of individual tasks on a non-partitioned single core processing element, $\delta_p$ is delay caused by partition switching, $\delta_c$ inter-processor communication delay.}
\label{Tab:comp}
\begin{tabular}{l|c|c|c}
\hline \hline
                                                   & Proposed architecture                                                                                      & Single Core Equivalent                                                                                                        & Fine-Grained Multithreaded                                                                             \\ \hline
Total WCET of independent tasks                    & $\tau_a + \tau_b + \tau_c + 2* \delta_p$ & $max(\tau_a, \tau_b,  \tau_c)  $                             & $3\times max(\tau_a, \tau_b,,\tau_c) $  \\ \hline
Total WCET of inter-dependent tasks               & $\tau_a + \tau_b + \tau_c + 2* \delta_p$ & $\tau_a + \tau_b + \tau_c + 2* \delta_c$ & $3\times(\tau_a + \tau_b + \tau_c)$ \\ \hline
WCET of an individual task, n               & $\tau_n$ & $\tau_n$ & $3\times \tau_n$ \\ \hline
Separate memory devices (instruction) & Not-required                                                                                                         & Required                                                                                                         & Required                                                                             \\ \hline
Separate memory devices (data)        & Not-required                                                                                                         & Required                                                                                                        & Required                                                                             \\ \hline
Efficient execution type                                       & fixed-time scheduled execution                                                                                  & parallel execution                                                                                         &  concurrent execution                                                    \\ \hline
\end{tabular}
\end{center}
\end{table*}
\renewcommand{\arraystretch}{1}
\subsection{{Use Case: Avionics }}
The objective of a use case analysis for the proposed system is not only to demonstrate the system performance against practical aerospace applications, but also to analyze the feasibility of the development of applications for the proposed architecture with conventional practice and following the guidelines in software development considerations i.e. DO-178B and DO-178C.

For demonstration, we have considered three applications for three partitions with different criticality-levels: 
\begin{itemize}
    \item \textit{A flight director (DAL A)}. 
    \item \textit{An autopilot with, altitude hold, auto-throttle, and heading-control (DAL B)}.
    \item \textit{A moving map application (DAL C)}.
\end{itemize}
The above applications are common for manned and unmanned airborne platforms, which can give a better understanding to the readers with different backgrounds.
A flight director (FD) application receives data from the \textit{flight data computer} and \textit{air data computer} about flight heading, altitude, attitude, airspeed etc. and integrate all the data into a command signal displayed on an attitude indicator. The flight director application can be used with or without autopilot. When used with autopilot, the FD can send reference commands to the autopilot. Without autopilot engaged, the FD presents all processed information as \textit{cues} to the pilot and the pilot needs to manually control the stick to follow the cues. 

The autopilot application, when engaged, maintains a reference altitude, speed and heading set by the pilot or the flight director. There are different ways of implementing the autopilot system. In our case, the autopilot can be engaged without engaging the flight director system. The autopilot system receives data from the \textit{flight data computer} and \textit{air data computer} about flights heading, altitude, attitude, airspeed etc. like flight director and computes control command for the control surfaces and the propulsion systems according to the preset reference states.

A moving map application displays the aircraft's location on a variety of pre-stored maps. For the experimentation, we have used 2D terrain map for the moving map application.

Each application controls functionality of a system, which is consist of several subsystems i.e. tasks in the context of software. For example, the autopilot system has multiple subsystems; vertical attitude control, lateral attitude control, velocity control, braking control etc. which are separate tasks. 

For software development, we have considered model-based development technique. Model-based development is a well adopted practice for software development in aerospace industries \cite{10.1007/978-3-540-79707-4_7}. In a model-based development approach, a mathematical model of the requirement is developed and latter C or C++ code is automatically generated by a tool, based on the model. Such tools provide traceability between the generated code and the requirements. In this work, we have used Simulink tool for model-based development for the proposed platform to generate standard C code. The code generated by Simulink can be certified by DO-178B standards when required guidelines are followed.  
The development process for applications of different criticality is kept separated. Each subsystem, that has a unique functionality, is developed as a function and called by the system it belongs to. Each partition may have multiple systems and subsystems under it. 

\subsubsection*{Setup}In this demonstration, we have used X-plane 11 flight simulator, a matured flight simulation software used in multiple pilot trainer simulator for simulating the dynamics of the airborne platform. The simulator features interfacing of external system to compute control commands to manipulate the actuators of the aircraft and the simulator simulates the dynamics of the aircraft based on the actuation. The X-plane 11 has an inbuilt UDP interface for sensor data transmission and control command reception. However, the \AE r\o \space platform does not feature UDP interface and an \textit{interfacing software}, that executes on the simulator host system, is used as a UDP-UART converter and vice-versa. 

\begin{figure}[h]
\begin{center}
\includegraphics[scale= 0.6]{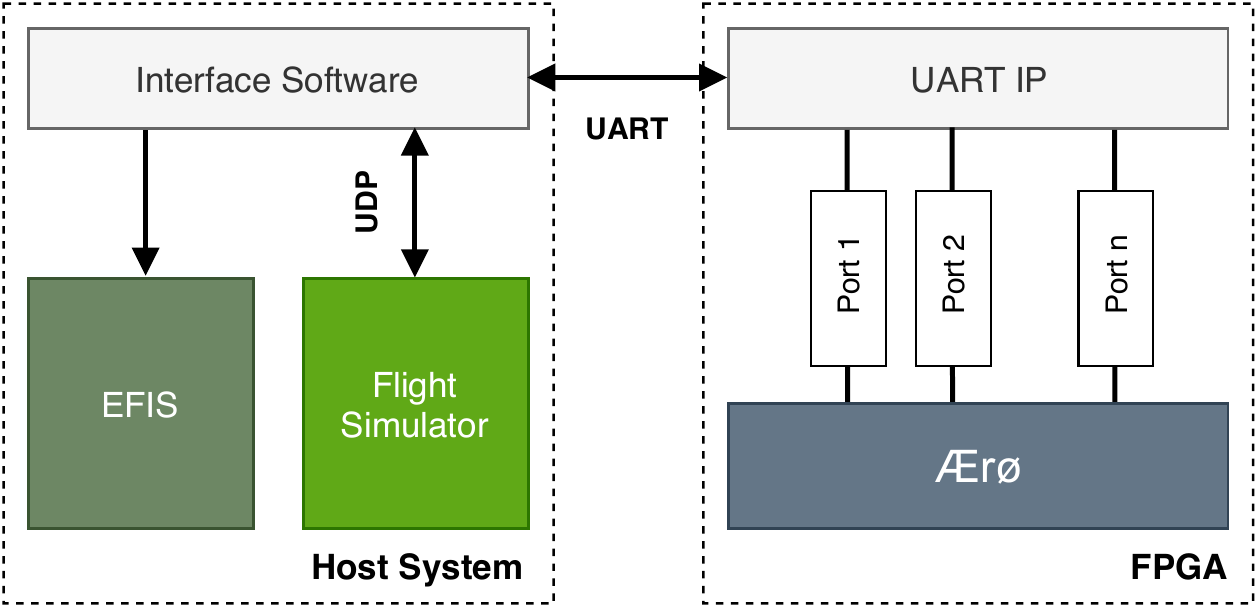}
\caption{The avionics use case demonstration setup.}
\label{fig:setup}
\end{center}
\end{figure} 
A generic UART module is not ideal for transmitting crucial data to a partitioned system, as reception buffer may contain data intended for one partition and get emptied by another partition. One way of dealing with this is having individual UART connections for each partition. However, we have used a custom UART IP to store the data packets in separate sampling ports. Each partition can read any of the sampling ports, but the port is only refreshed (overwritten) when a new sample arrives.
Furthermore, the control commands received from the \AE r\o \space platform for FD and moving map applications are fed into a software defined electronic flight instrument system (EFIS) for visual inspection for correctness. A block diagram of the demonstration setup is presented in Figure \ref{fig:setup}.

\vspace{2mm}

The autopilot is required to compute control commands in a cyclic loop and the required control loop frequency depends on the dynamics of the airborne platform and has a typical value of 50Hz for large airliners to 500Hz for agile unmanned systems. In this experiment, we have considered the execution frequency of the autopilot application to be 200 Hz, which is our limitation due to bandwidth constraints between the simulator and the platform. Additionally, the flight director application should have the same execution frequency to feed reference signals to the autopilot system. A moving map does not need a high execution frequency and a frequency of 10Hz has been considered.

The Table \ref{tab:sim} shows the WCET of each application and the execution time assigned to each partition. Note that the partition execution time always is higher than the execution time of the application running within the partition. 

\renewcommand{\arraystretch}{1.5}
\begin{table}[h]
\begin{center}

\caption{Timing analysis of applications and partitions}
\label{tab:sim}
\begin{tabular}{lcc}
\hline
\hline
\textit{Application}                 & \begin{tabular}[c]{@{}c@{}}App. WCET ($\tau_a$) \\ (ms)\end{tabular} &\begin{tabular}[c]{@{}c@{}}Partition WCET ($\tau_p$) \\  (ms)\end{tabular} \\ \hline
{Autopilot System}       &        1.003        &         2               \\ \hline
{Flight Director  \space \space \space \space} &       1.127         &        2              \\ \hline
{Moving map }      &        0.319        &         1           \\ \hline
\end{tabular}
\end{center}

\end{table}
\renewcommand{\arraystretch}{1}

\begin{figure}[h]
\begin{center}
\includegraphics[scale= 0.36]{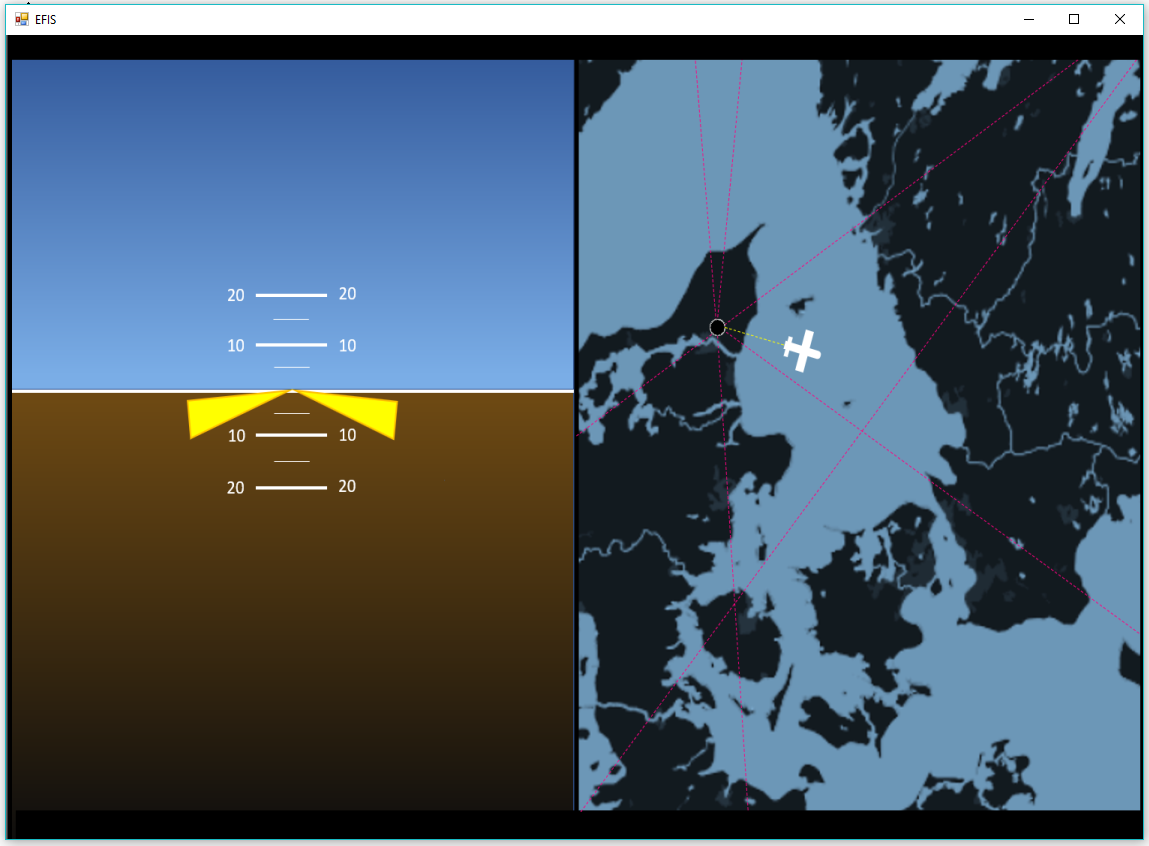}
\caption{Showing a screenshot of software driven EFIS application on the host system. The FD cues and moving map applications are driven by the control command received from FD and moving-map application, executing on \AE r\o \space platform.}
\label{fig:efis}
\end{center}
\end{figure}

A simulated flight test is conducted in the simulated environment where a test flight is flown for more than 10 Hrs in a pre-defined flight path by the autopilot application driven by the FD application, both executing on the \AE r\o \space platform. For graphical output, the FD cues driven by the flight director application is displayed over an electronic attitude director indicator (EADI) on a primary flight display (PFD) system and a screenshot is shown in Figure \ref{fig:efis}.

\section{Related Work}
The concept of moving RTOS operation in hardware is not new. In FASTCHART \cite{144077}, deterministic execution was achieved with 64 different tasks with 8 different priorities with a RISC architecture by removing cache and pipeline. Adomat et al. in \cite{557849} presented a real-time hardware kernel that can support a similar number of tasks and priorities like FASTCHART, but with additional hardware features.

In recent years, Zimmer et al. \cite{6925994} has introduced a fine-grained multithreaded processor architecture, FlexPRET, for mixed-criticality systems. The work demonstrates a WCET analyzable framework that accommodates isolation in both temporal and spatial domain without wasting any computational cycle. Tasks are segregated in threads and each thread is given access to the computational resources for a single clock cycle in fixed or active-round-robin arbitration. 
A very similar architecture PTARM \cite{6378622} by Liu et al., is a fine-grained multithreaded architecture with fixed-round-robin arbitration among four threads. The architecture avails hard-real-time execution time, however, cycles are wasted if all the threads are not active. 
Another similar architecture XMOS \cite{6341002} poses better utilization of resources by excluding inactive tasks from the scheduling. WCET is analyzed by considering the maximum number of active tasks at any given time. Ungerer et al. \cite{5567091} introduced a worst-case time-analyzable multi-core architecture for mixed-criticality system with isolation between tasks. Although, the architecture is limited to a single hard-real-time task per core. 
Delvai et al. \cite{1212740} introduced a 16-bit, 3 stage processor, SPEAR, with repeatable-time instructions by single path execution flow.

\section{Conclusion and Future Work}
Virtualization is an essential requirement for mixed-criticality systems to mitigate inter-partition interference. Hardware-based partitioning can provide an adequate solution for virtualization without any software support. For cost-effective platforms with SWaP constraints, such a system can be beneficial as it saves the cost and resources that would be spent on a software-based partitioning mechanism otherwise. Furthermore, a hardware-scheduling system can provide cycle-accurate partitioning switching to accommodate hard-real-time requirements. 

In this research, we have considered a non-preemptive fixed-scheduling arbitration in the hardware scheduler for cycle-accurate timing analysis for possible implementations in airborne systems. There is an opportunity to improve the scheduling algorithm to support dynamic and priority-based arbitration. 
Furthermore, the proposed architecture can possibly be extended to multicore implementation. By interfacing with on-chip interconnects, the proposed processor core can be used for single core equivalent multicore platform. Lastly, the platform can be demonstrated on a physical testbed for unmanned flights.


\bibliography{bibfile}

\begin{IEEEbiography}[{\includegraphics[width=1in,height=1.25in,clip,keepaspectratio]{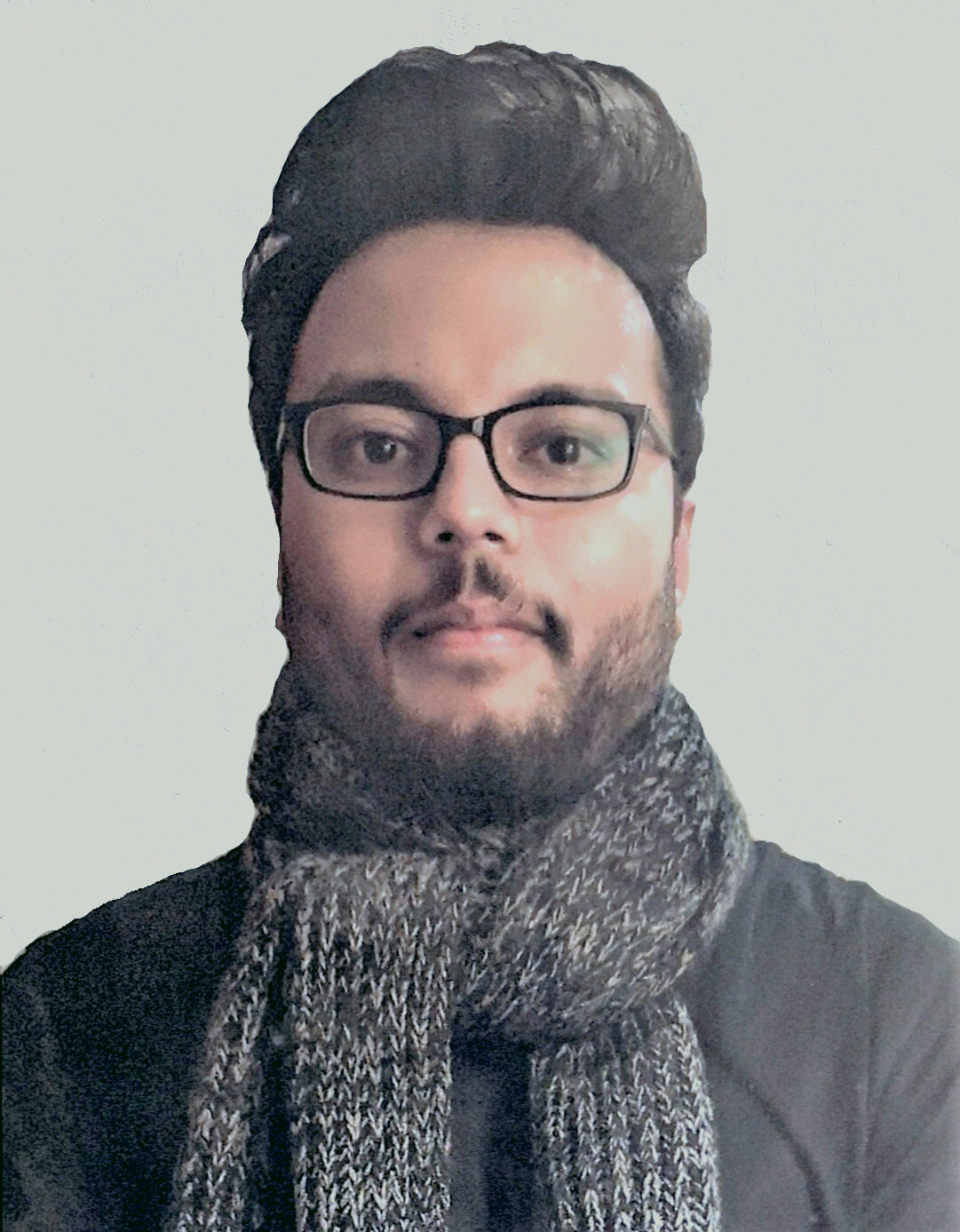}}]{Shibarchi Majumder}is employed as a research fellow at the Department of Electronic Systems, Aalborg University, Aalborg, Denmark. 
He has years of industrial and academic experience in avionics systems, embedded flight computing, safety-critical systems and has several publications in related domains. 

His research interests include real-time systems, mixed-criticality systems, safety-critical systems, hardware design, embedded computation and unmanned aerial systems. Earlier, he
received a bachelor's degree in Aerospace Engineering and a master's degree in Avionics Engineering.
\end{IEEEbiography}
\vskip -2\baselineskip plus -1fil
\begin{IEEEbiography}[{\includegraphics[width=1in,height=1.25in,clip,keepaspectratio]{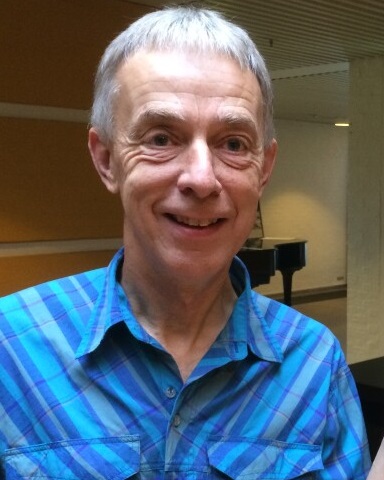}}]{Jens Frederik Dalsgaard Nielsen}
is employed as Associate Professor at Aalborg University at the section Automation \& Control. He has a Master of Science in EE and a PhD within automation and control domain.
 
For more than 15 years he has been heading the student satellite activities at Aalborg University which has launched 5 cubesats 100\% developed at AAU and participated in three other launches. 
His primary domain is realtime systems ranging from hardware to real time operating systems, networking for safety critical systems and software development.
\end{IEEEbiography}

\vskip -2\baselineskip plus -1fil
\begin{IEEEbiography}[{\includegraphics[width=1in,height=1.25in,clip,keepaspectratio]{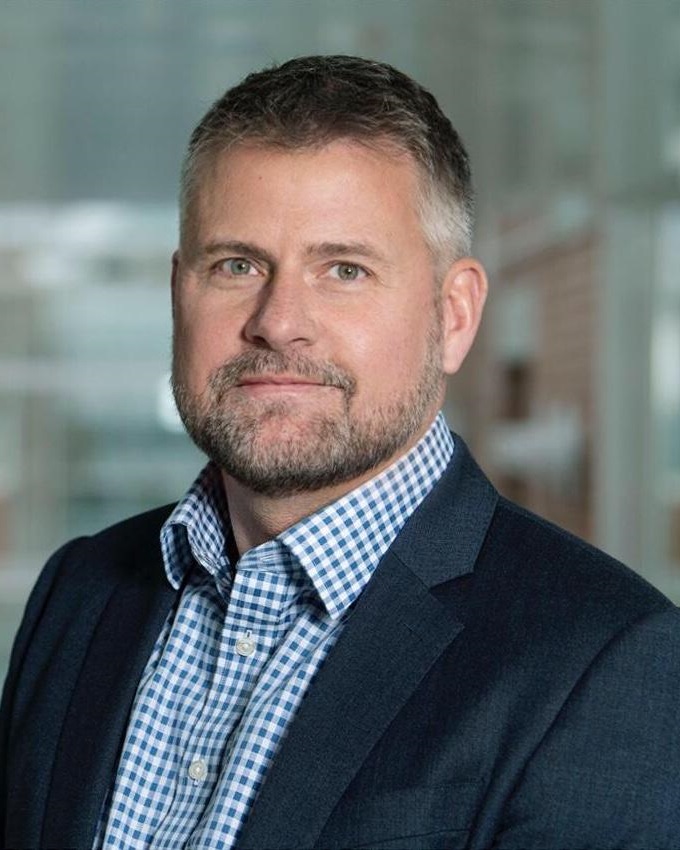}}]{Thomas Bak}
received a PhD degree in control systems from Aalborg University in 1998. He became an Assistant Professor in 1998, an Associate Professor in 1999, and a Full Professor of autonomous systems in 2006. From 2003-2006 he was a senior researcher and head of research unit at Aarhus University. Since 2018 he is head of the Department of Electronic Systems. He has published more than 100 papers in the fields of control and its applications. His current research interest includes autonomous systems and robotics. He is a senior member of IEEE and chair of the IEEE Joint Chapter on Control System Society and Robotics and Automation Society. He is associated editor for the European Journal of Control.
\end{IEEEbiography}





\end{document}